\documentclass[utf8]{FrontiersinHarvard} 

\usepackage{url,hyperref,lineno,microtype,subcaption}
\usepackage[onehalfspacing]{setspace}
\usepackage[T1]{fontenc}

\def\keyFont{\fontsize{8}{11}\helveticabold }
\def\firstAuthorLast{Y.Q. Zhao {et~al.}} 
\def\Authors{Yuqian Zhao\,$^{1,2}$, Bing Sun\,$^{3}$, Zhan-Feng Mai$^{4}$ and Zhoujian Cao\,$^{1,2,*}$}
\def\Address{$^{1}$Institute for Frontiers in Astronomy and Astrophysics, Beijing Normal University, Beijing 102206, China \\
$^{2}$Department of Astronomy, Beijing Normal University, Beijing 100875, China\\
$^{3}$CAS Key Laboratory of Theoretical Physics, Institute of Theoretical Physics, Chinese Academy of Sciences, Beijing 100190, China \\
$^{4}$ Kavli Institute for Astronomy and Astrophysics, Peking University, Beijing 100871, China}
\def\corrAuthor{Zhoujian Cao}

\def\corrEmail{zjcao@amt.ac.cn}

\begin{document}
\onecolumn
\firstpage{1}

\title[QNMs and Detection during Ringdown]{Quasi Normal Modes of Black Holes and Detection in Ringdown Process} 

\author[\firstAuthorLast ]{\Authors} 
\address{} 
\correspondance{} 

\extraAuth{}

\maketitle

\begin{abstract}

Quasi-normal modes (QNMs) of a black hole (BH) are the eigen modes describing the dissipative oscillation of various fields in that spacetime, which can be intrinsically produced by the linear perturbation theory. With the discovery of the first gravitational waves (GWs) event, GW150914, a new window into the universe has been opened, allowing for the detection of QNMs associated to the ringdown process, which will enable more accurate measurements of the BHs parameters as well as further testing of general relativity. This article discusses the linear perturbation theory of BHs and provides review of several QNMs calculation methods including the newly developed methods. We will also focus on the connection between QNMs and the detection of GWs as well as some recent advancements in this area.


\tiny
 \keyFont{ \section{Keywords:} Quasi normal modes (QNMs), Gravitational waves (GWs), Ringdown, Black holes (BHs), General relativity (GR)} 
\end{abstract}

\section{Introduction}

    Black holes (BHs);~\cite{einstein1915sitzungsberichte} and gravitational waves (GWs);~\cite{einstein1916approximative} are proposed in the 20th century as the significant ingredients predicted by general relativity (GR). In the next several decades, people are attempt to find out and explore their existent evidences and characteristics;~\cite{Barack_2019}. Since 2015, several GWs events, such as GW150914;\cite{GW150914} and GW170817A;~\cite{GW170817A}, from the binary black holes (BBHs) of the stellar mass and the binary Neutron stars (BNSs) respectively, have been detected;~\cite{GWOSC,GWTC1,GWTC2,GWTC2.1,GWTC3}, while the optical observations for the shadow from M87 also provided further indirect evidence for supermassive BHs (SMBHs);~\cite{collaboration2019first}.
    
    BHs own a natural distinguishable feature --- an event horizon. This surface, as a one-way causal boundary, separates the communication of information in a classical level and brings a huge obstacle for us to observe the interior;~\cite{1916skpa.conf..424S}. Therefore, the BHs, which are the vacuum solutions can only determined by several parameters without the complicated equations of state like stellar or neutron stars;~\cite{kokkotas1999quasi,nollert1999quasinormal}, such as Schwarzschild and Kerr BHs, are intrinsically among the simplest objects in the GR or other metric gravity theory. However, practically, there almost does not exist an isolated black hole because the astronomical environment like dark matter spike and accretion disk near a BH is complex and changeable;~\cite{Nampalliwar_2021,Xu_2021}. Nevertheless, compared with the astrophysical BHs with huge mass, the objects around these BHs own less mass. Interacting with these various objects, the BH is thus perturbed. As one of observable evidences for such perturbation, the GWs then will be generated and reach our detector in the solar system;~\cite{GWreview2}. Note that the ringdown phase of a binary system could be depicted in a similar scenario. Exploring the BH perturbation theory naturally gives rise to the topic of quasi-normal modes (QNMs).
    
    QNMs of BHs as the eigen modes describing the dissipative oscillation of various fields in corresponding perturbed spacetime;\cite{berti2009quasinormal}, have been discussed for decades since it was initially proposed by Regge and Wheeler during the analysis of the stability of a Schwarzschild BH;~\cite{RW}. Specifically, at linear perturbation level, the perturbed metric results in a set of homogeneous second order differential equations with the discrete complex eigen value $\omega$ called QNMs frequency, only if we set the incoming boundary behavior near horizon and outgoing at spatial infinity with more detailed definition in Sec.\ref{section:QNMeqn}.In addition to the stability of BHs, the QNMs usually produce GWs with the combination of discrete modes in the frequency domain corresponding to the ringdown stage of a binary BH merger GW event, and uniquely is determined by the parameters of the BH, similar to the spectrum of hydrogen atoms in quantum mechanics or the "sound" of BHs, they are therefore also called the BHs spectroscopy or overtone~\cite{berti2006gravitational,nollert1999quasinormal}. The precise measurements of such a GW signal allow us to precisely determine the parameters of BHs and test no-hair theorem or further GR properties;~\cite{PhysRevLett.123.111102,abbott2021tests}. After decades of development, the QNMs has been extented from the original stability analysis to include properties itself, calculating methods, the GWs, etc.
    
    The high precision measurement also requires accurately calculating the QNMs. Methods for the exact calculation of QNMs have also been developed for decades, and the most commonly used are the WKB method and the continued fraction method;~\cite{schutz1985black,leaver1985analytic}. Thoughts on the difficulties of calculation of QNMs and some methods will be reviewed in Sec.\ref{section:Method}. In general, the linear perturbation theory is also sometimes used for exploring the generation of GWs in other cases such as the extreme mass ratio inspirals;~\cite{PhysRevD.102.024041}, where the inhomogeneous equations are introduced with the Green function method;~\cite{poisson2011motion}. The latter problem will not be involved, however, due to the natural association between homogeneous and inhomogeneous solutions through Green functions;~\cite{PhysRevD.34.384,10.1143/PTPS.128.1}, some of the applications for latter case will also be mentioned.

    With the discovery of the first GW event GW150914;~\cite{GW150914}, who opened a new window into the universe, the data-driven exploration of BHs is now possible, allowing for accurate measurements of the BHs parameters as well as additional testing of general relativity;~\cite{GWReview1}. However, the remnant of BHs just after merge stage is highly nonlinear which is not applicable for linear perturbation theory, thus how to connect ringdown waveform determined by QNMs is still under discussion as reviewed in Sec.\ref{Section:detection}. Generally speaking, due to the no-hair theorem, we believe that the properties of black holes are determined by mass of BH $M$, spin $a=\frac{J}{M}$ with $J$ the angular momentum, and charge $Q$;~\cite{penrose1969gravitational,PhysRevLett.26.331,hansen1974multipole,PhysRevLett.114.151102}. However, it has been shown that the charge of a BH have no detectable effect on the ringdown waveform;~\cite{PhysRevD.105.062009}, hence, the charged BHs will not be involved while we discuss the rotating BHs of Kerr case.
    
    In this article, we will review the theory of QNMs with their contribution to the ringdown stage of the GWs events. There have been several excellent reviews for QNMs;~\cite{nollert1999quasinormal,kokkotas1999quasi,berti2009quasinormal,konoplya2011quasinormal}. These articles provide a comprehensive exploration of this topic from the aspects of theory, method, detection, etc. However, with the development of the past decade, there have been significant advancements in all aspects:
    
    Initially, QNMs equations in more different spacetime for tensor cases associated to GWs were explored, while the reconstruction of metric was developed. In the meanwhile, further methods for calculating QNMs or related eigen functions have been proposed with several available programs. Furthermore, with the successful detection of GWs events;~\cite{GW150914}, the analyses for the detection of GWs during the ringdown process are now under discussion. We will focus on these advancements and review these aspects as follows:
    
    We review the QNMs produced by linear perturbation theory in Sec.\ref{section:QNMeqn} and the method to calculate the QNMs in Sec.\ref{section:Method}. At the end of the article in Sec.\ref{Section:detection}, we review the detection advancements based on the detected data from LIGO;~\cite{LIGO}, VIRGO;~\cite{VIRGO} and KAGRA;~\cite{KAGRA}. Without further explication, we'll use the units $\hbar=c=G=1$.

\section{Linear Perturbation Theory and Master Equations of QNMs}\label{section:QNMeqn}

    In general, the perturbations of a BH can result from either an additional field injected into spacetime (such as a particle with the mass $m\ll M_{BH}$ falling into a BH;~\cite{SchFallingParticle}) or directly from the perturbing metric (such as the ringdown process at the end of a binary BH merger event). Within general relativity (as well as several other gravity theories), the linear perturbation theory requests us to focus on the first order of perturbation and ignore the reaction to the background. Fields or gravitational radiation will propagate in spacetime in the form of damping oscillations, the characteristics of which are typically governed by a set of radial Schrödinger-like equations in frequency domain with the corresponding angular equations
    
    There are two methods to study the linear perturbations in the background spacetime: Firstly, one can parameterize the perturbations as the variation of the coefficients of metric directly and insert them into the Einstein equation or Maxwell equation in the curved spacetime. Or, we can also study them in the form of Newman-Penrose (N-P) formalism;~\cite{NP} via the N-P equations. Both of the two methods in several cases has been summarizes in the monograph from Chandrasekhar;~\cite{chandrasekhar1985mathematical}.
    
    In this section, we give a general review of the Schrödinger-like equations that govern a BH's quasi-normal modes, specifically for Schwarzschild and Kerr BHs. The master equations governing the propagation of fields or gravitational radiation will be our starting point for discussion.
    
\begin{itemize}
     \item\textbf{Scalar Field in Background Spacetime (Scalar Perturbations).} The motion of a massless scalar field $\Phi$ in the background spacetime can be obtained from Klein-Gordon equation:
        \begin{equation}
            \nabla^\mu \nabla_\mu\Phi=0
        \end{equation}
    where $\nabla_\mu$ is the covariant derivative. The equation mentioned above may be formally rewritten as:
    \begin{equation}\label{CGeqn}
        \frac{1}{\sqrt{-g}} \partial_\mu\left(\sqrt{-g} g^{\mu \nu}  \partial_\nu \Phi\right)=0
    \end{equation}
    with $g$ the determinant of the background metric $g_{\mu\nu}$.
    
    \item\textbf{Maxwell Field in Background Spacetime (Vector Perturbations).} In this case, the Maxwell equations govern the propagation of a massless vector field $A_\mu$ in background spacetime:
    \begin{equation}
    \nabla^\mu F_{\mu \nu}=0,\quad\text{with}\quad F_{\mu \nu}=\partial_\mu A_\nu-\partial_\nu A_\mu 
    \end{equation}
    which can be rewritten in a more explicit form in the curved background spacetime as:
    \begin{equation}\label{Maxwelleqn}
    \partial_\nu\left[\left(\partial_\alpha A_\sigma-\partial_\sigma A_\alpha\right) g^{\alpha \mu} g^{\sigma \nu} \sqrt{-g}\right]=0
    \end{equation}
    
    \item\textbf{Linear Gravitational Perturbation in Background Spacetime (Tensor Perturbations).} Within a gravity theory, "gravitational perturbation" refers to the perturbation of spacetime itself. For metric perturbations, the metric can be expressed explicitly as:
    \begin{equation}\label{LinearMetric}
    g_{\mu \nu}=\mathring{g}_{\mu \nu}+\delta g_{\mu \nu}+O(\delta g^2_{\mu\nu})
    \end{equation}
    where $\mathring{g}_{\mu \nu}$ is the metric of background spacetime, $\delta g_{\mu \nu}$ is the linear perturbation term, the terms of two order perturbations $g_{\mu \nu}^2$ and higher orders are disregarded due of their little impact in comparison to $\delta g_{\mu \nu}$. And the governing equations are well known Einstein equations provided by:
    \begin{equation}\label{EinsteinEqn}
    G_{\mu\nu}=R_{\mu \nu}-\frac{1}{2} R g_{\mu \nu}=0
    \end{equation}

\end{itemize}

\begin{figure}[h!]
    \begin{center}
    \includegraphics[width=10cm]{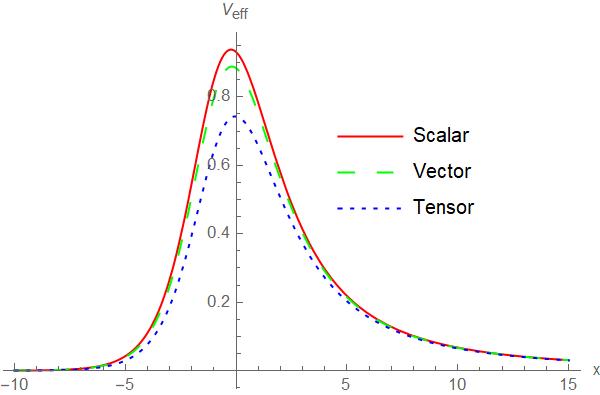}
    \end{center}
    \caption{This figure shows the tendency of effective potentials asymptotically approach to 0 at horizon and spatial infinity, where we choose the Schwarzschild case with $\ell=2$ and $c=G=2M=1$ as defined in Eq.\eqref{SchEffV}}\label{fig:SchEffV}
\end{figure}

After applying the separation of variables to a decoupled master equation, the radial Schrödinger-like equations can be obtained in the form of:
    \begin{equation}\label{QNMeqn}
        \left(\frac{\mathrm{d}^2}{\mathrm{d} x^2}+(\omega^2-V_{\text{eff}})\right) \Psi= 0
    \end{equation}
    where $\omega$ is the eigen frequency coming from the separation of $t$ under the Fourier transform, and $V_{\text{eff}}$ is the effective potential determined by the background metric with in general the asymptotic boundary behaviors at infinity as (for example, we show the effective potential in Schwarzschild case in Figure.\ref{fig:SchEffV}):
    \begin{equation}
        V_{\text{eff}}=\begin{cases}
        0 & x\rightarrow-\infty  \\
            0 & x\rightarrow+\infty  
            \\\end{cases}
    \end{equation}
    leading to the asymptotic boundary behaviors of eigen function $\Psi$ at infinity determined by the solutions of:
    \begin{equation}\label{eqBCs}
        \left(\frac{\mathrm{d}^2}{\mathrm{d} x^2}+\omega^2 \right) \Psi= 0,\qquad x\rightarrow \pm \infty
    \end{equation}
    The general solutions near boundaries can be written in the combination of $\mathrm{e}^{+\mathrm{i} \omega x}$ and $\mathrm{e}^{-\mathrm{i} \omega x}$ describing \emph{outgoing} and \emph{incoming waves} respectively. And the different combination of the both solutions usually results in three different directions:
    
    \begin{itemize}
    
    \item\textbf{$\Psi \rightarrow \mathrm{e}^{-\mathrm{i}\omega x}$ (incoming) at $-\infty$ and $\Psi \rightarrow \mathrm{e}^{+\mathrm{i}\omega x}$ (outgoing) at $+\infty$}. This will result in the most fundamental field of \emph{quasi-normal modes (QNMs)}, which is the subject of this article. In this case the imaginary component of $\omega$ is often negative due to the stability;~\cite{RW}.
    
    \item\textbf{$\Psi \rightarrow \mathrm{e}^{-\mathrm{i}\omega x}$ (incoming) at $-\infty$ and $\Psi \rightarrow \mathrm{e}^{-k x}$  at $x \to +\infty$}. This will give rise to a new discipline referred to as \emph{quasi-bound states (QBS)} after considering the mass of fields. $k \equiv \sqrt{m_p^2 - \omega^2}$ where $m_p$ denotes the mass of the massive scalar perturbation. The QBS's often employed to explore \emph{superradiant instability} ~\cite{brito2020superradiance}, which is not involved in this article.
    
    \item\textbf{$\Psi \rightarrow A\mathrm{e}^{-\mathrm{i}\omega x}+B\mathrm{e}^{\mathrm{i}\omega x}$ at $-\infty$ and $\Psi \rightarrow \mathrm{e}^{-\mathrm{i}\omega x}$ (outing) at $+\infty$ with $B\neq0$}. This extends a series of studies on \emph{exotic compact objects (ECOs)};~\cite{cardoso2019distinguishing,maggio2021extreme,sago2021oscillations,cardoso2019testing}, which will not be taken in our consideration.
    
    \end{itemize}

    For scalar case, there is just one component equation, which is inherently decoupled. However, it is challenging to obtain a decoupled equation in other cases since a Maxwell field is governed by six coupled component equations (Eq.\eqref{Maxwelleqn}) whereas gravitational perturbations are governed by ten (Eq.\eqref{EinsteinEqn}). Actually, avenue to the decoupled equation must take the symmetry of background spacetime corresponding to the gauge-invariant variables into account with expressing the master equation in terms of them;~\cite{kodama2000brane}, or N-P formalism may be also helpful;~\cite{NP}. We will discuss how to address this problem in some situations, properly speaking in the Schwarzschild and Kerr cases. At the end of this section, we also summarize several publications including the QNMs equations for tensor perturbation case.
        
    \subsection{Perturbations in Schwarzschild Spacetime}
    
    With \textbf{the effect of a scalar field in vacuum}, we start our discussion in Schwarzschild spacetime. The static spherical BH solution is the well known Schwarzschild metric given by:
    \begin{equation}\label{SchMetric}
        \mathrm{d} s^2=-\left(1-\frac{2 M}{r}\right) \mathrm{d} t^2+\left(1-\frac{2 M}{r}\right)^{-1} \mathrm{~d} r^2+r^2 \mathrm{~d} \theta^2+r^2 \sin ^2 \theta \mathrm{d} \varphi^2
    \end{equation}
    where $M$ is the mass of BH. As is mentioned above, Eq.\eqref{CGeqn} is an inherently decoupled equation with $\Phi$ now the function of $(t,r,\theta,\phi)$. Similar to the technique taken to solve the hydrogen atom problem in quantum mechanics, by substituting the background metric provided by Eq.\eqref{SchMetric} into the master equation and applying the separation of variables:
    \begin{equation}
        \Phi(t, r, \theta, \phi)=\sum_{\ell=0}^{\infty}\sum_{m=-\ell}^{m=\ell}Y_{\ell m}(\theta, \phi) R (t, r) / r
    \end{equation}
    the QNMs governing equation can be obtained in time domain:
    \begin{equation}
        \left(\frac{\mathrm{d}^2}{\mathrm{d} r_*^2}-\frac{\mathrm{d}^2}{\mathrm{d} t^2}-V_{\mathrm{scalar}} \right) R(t,r) = 0
    \end{equation}
    where the effective potential is given by:
    \begin{equation}\label{SchPontential}
        V_{\mathrm{scalar}}(r)=(1-\frac{2M}{r})\left(\frac{\ell(\ell+1)}{r^2}+\frac{2M(1-s^2)}{r^3}\right),\quad \text{with}\; s=0 
    \end{equation}
    where $\ell(\ell+1)$ is the angular separation constant arising from the separation of angular part and the angular equations are:
    \begin{equation}\label{aSch}
        \gamma^{c d} \nabla_d \nabla_c Y_{\ell m}(\theta,\phi)=-\ell(\ell+1) Y_{\ell m}(\theta,\phi)
    \end{equation}
    with $\gamma=\mathrm{diag}(1,\sin^2\theta)$ the metric of the unit sphere surface $S^2(\theta,\phi)$ and $Y_{\ell m}(\theta,\phi)$ the scalar spherical harmonics.
    
    Suppose that the solution of the perturbation equation with the time dependence $R(t,r)=\mathrm{e}^{\mathrm{i}\omega t}R(\omega,r)$ corresponding to the Fourier transformation in the standard procedure of the normal modes analysis;~\cite{nollert1999quasinormal,kokkotas1999quasi}, and substituting it into Eq.\eqref{QNMeqn} yields the radial QNMs equations:
    \begin{equation}
        \left(\frac{\mathrm{d}^2}{\mathrm{d} r_*^2}+(\omega^2-V_{\mathrm{scalar}})\right) R= 0
    \end{equation}
    with $r_*$ the tortoise coordinate defined by:
    \begin{equation}\label{torto}
        \frac{\mathrm{d}r_*}{\mathrm{d} r}= \left(1-\frac{2M}{r} \right)^{-1}
    \end{equation}
    
   \textbf{The gravitational perturbation in Schwarzschild spacetime} is then taken into consideration, beginning with the symmetry of the Schwarzschild geometry corresponding to gauge-invariant variables;~\cite{thompson2017gauge,nagar2005gauge,martel2005gravitational,sarbach2001gauge}. Because of the static spherical symmetry of the background manifold $\mathcal{M}^4(t,r,\theta,\phi)$, it can be regard as the product of a Lorentzian 2-dimension manifold $M^2(t,r)$ and a 2-dimension unit sphere surface manifold $S^2(\theta,\phi)$  with the metric $\gamma=\mathrm{diag}(1,\sin^2\theta)$ as is mentioned in the scalar case. By taking advantage of this, the metric perturbations $h_{\mu\nu} \equiv \delta g_{\mu\nu}$ can be decomposed in multipoles known as odd-parity or even-parity depending on their transformation features under parity. The definitions of both multipoles are as follows: under a parity transformation $(\theta,\phi) \rightarrow (\pi - \theta, \pi + \phi)$, the \emph{odd} (or \emph{axial}) parity part transforms as $(-1)^{\ell+1}$ while the \emph{even} (or \emph{polar}) parity transforms as $(-1)^{\ell+1}$. Thus, the metric perturbations $h_{\mu\nu}$ in Eq.\eqref{LinearMetric} can be expressed as:
    \begin{equation}
        \delta g_{\mu \nu}=\sum_{\ell=0}^{\infty}\sum_{m=-\ell}^{m=\ell}\left[\left(\delta g_{\mu \nu}^{\ell m}\right)^{(\mathrm{odd})}+\left(\delta g_{\mu \nu}^{\ell m}\right)^{(\mathrm{even})}\right]
    \end{equation}
    
    Similar to the scalar case, the following step involves the separation of the variables. To do this, it is necessary to introduce the vector spherical harmonics;~\cite{nollert1999quasinormal,dewitt1973black,edmonds1996angular} and the tensor spherical harmonics;~\cite{thompson2017gauge,nagar2005gauge,thorne1980multipole,zerilli1970gravitational,mathews1962gravitational,RW} as the angular dependence. And the perturbation of the odd parity could be described as:
    \begin{equation}
       \left(\delta g_{\mu \nu}^{\ell m}\right)^{(\mathrm{odd})}=\left[\begin{array}{cccc}
        0 & 0 & 0 & h_0(r) \\
        0 & 0 & 0 & h_1(r) \\
        0 & 0 & 0 & 0 \\
        h_0(r) & h_1(r) & 0 & 0
        \end{array}\right]\left(\sin \theta \frac{\partial}{\partial \theta}\right) Y_{l 0}(\theta)\mathrm{e}^{\mathrm{i}\omega t}
    \end{equation}
    while that of the even parity is:
    \begin{equation}
        \left(\delta g_{\mu \nu}^{\ell m}\right)^{(\mathrm{even})}=\left[\begin{array}{cccc}
        H_0(r) (1-\frac{2M}{r}) & H_1(r) & 0 & 0 \\
        H_1(r) & H_2(r) (1-\frac{2M}{r})^{-1} & 0 & 0 \\
        0 & 0 & r^2 K(r) & 0 \\
        0 & 0 & 0 & r^2 K(r) \sin ^2 \theta
        \end{array}\right] Y_{l 0}(\theta)\mathrm{e}^{\mathrm{i}\omega t}
    \end{equation}
   with $h_0$, $h_1$, $H_0$, $H_1$, $H_2$ and $K$ the parameterization coefficients of the perturbation metric as for the function of $r$, and the Fourier transformation has been employed. The above formalism is derived from the work of \cite{vishveshwara1970stability} by using spherical symmetry resulting to $m=0$ under the Regge-Wheeler gauge;~\cite{kokkotas1999quasi,nollert1999quasinormal,nagar2005gauge,RW,thompson2017gauge}. After substituting the parameterized metric formalism into the Einstein equation provided in Eq.\eqref{EinsteinEqn}, ten coupled two-order differential equations governing gravitational perturbations will be obtained, with three for odd parity and seven for even;~\cite{berti2009quasinormal}. 
   
   In order to obtain a decoupled master equation, specific parameterization coefficient combinations must be introduced. However finding the specific combination is challenging, fortunately that of odd parity was first found in \cite{RW} with some mistakes, and rectified by \cite{RWCorrect}. Then the decoupled QNMs equation in the form of Eq.\eqref{QNMeqn} can be obtained:
    \begin{equation}
        \left(\frac{\mathrm{d}^2}{\mathrm{d} r_*^2}+(\omega^2-V_{\mathrm{tensor}})\right) R= 0
    \end{equation}
   where $r_*$ is the defined by Eq.\eqref{torto} and the effective potential is given by:
    \begin{equation}
        V^{(\mathrm{odd})}_{\mathrm{tensor}}(r)=\left(1-\frac{2 M}{r}\right)\left[\frac{\ell(\ell+1)}{r^2}+\frac{2 M\left(1-s^2\right)}{r^3}\right],\quad \text{with}\; s=2
    \end{equation}
    which is identical to the scalar case in Eq.\eqref{SchPontential} except of the value of $s$. The QNMs governing equation above for the odd parity called \emph{Regge-Wheeler equation}.
    
    Meanwhile there is also a decoupled equation for even parity with the effective potential given by:
    \begin{equation}
        V^{(\mathrm{even})}_{\mathrm{tensor}}(r)=\left(\frac{1-2 M}{r}\right) \frac{2 \Lambda^2(\Lambda+1) r^3+6 \Lambda^2 M r^2+18 \Lambda M^2 r+18 M^3}{r^3(\Lambda r+3 M)^2}
    \end{equation}
    with $\Lambda=\frac{1}{2}(\ell-1)(\ell+2)$ and the corresponding governing equation for even parity are called \emph{Zerilli equation} originally derived by \cite{zerilli1970effective,zerilli1970gravitational}, with the corrected version can be found in the Appendix A of \cite{sago2003gauge}.
    
    After that, by introducing gauge-invariant variables, which were initially proposed by \cite{moncrief1974gravitational}, a set of normative and efficient procedures for gravitational perturbations in Schwarzschild spacetime was constructed from the following research can be found in \cite{thompson2017gauge,nagar2005gauge,martel2005gravitational,sarbach2001gauge,gerlach1979gauge,gerlach1980gauge}.

    The research for both of the multipoles above yields a significant property called \emph{isospectral} first discovered by Chandrasekhar in his book;~\cite{chandrasekhar1985mathematical} with some discussion can be found in Appendix A of \cite{berti2009quasinormal} and the recent research in \cite{jaramillo2022gravitational}, indicating that various multipoles may generate the same characteristic spectrum. That implies it is sufficient to analyze either of the situation for simplification.

    Along the same avenue as before, \textbf{the Maxwell field in Schwarzschild spacetime} can be considered by expressing the Maxwell equations Eq.\eqref{Maxwelleqn} into the vector harmonics and decoupled into the QNMs equation in the form of Eq.\eqref{QNMeqn}:
    \begin{equation}
        \left(\frac{\mathrm{d}^2}{\mathrm{d} r_*^2}+(\omega^2-V_{\mathrm{vector}})\right)R= 0
    \end{equation}
   where $r_*$ has the same definition as Eq.\eqref{torto} and the effective potential is given by:
    \begin{equation}
        V_{\mathrm{vector}}(r)=\left(1-\frac{2 M}{r}\right)\left[\frac{\ell(\ell+1)}{r^2}+\frac{2 M\left(1-s^2\right)}{r^3}\right],\quad \text{with}\; s=1
    \end{equation}
    
    \textbf{For a concise summary}, the QNMs equation in Schwarzschild spacetime is provided by:
    \begin{equation}\label{Scheqn}
        [\frac{\mathrm{d}^2}{\mathrm{d} r_*^2}+(\omega^2-V_{\mathrm{Sch}})] R= 0
    \end{equation}
     with $r_*$ the tortoise coordinate defined the same as Eq.\eqref{torto}:
    \begin{equation}
        \frac{\mathrm{d}r_*}{\mathrm{d} r}= (1-\frac{2M}{r})^{-1}
    \end{equation}
    who maps $r$ from the region $(2M,+\infty)$ to the region $(-\infty,+\infty)$ with $2M$ the horizon of Schwarzschild, and the effective potentials are:
    \begin{equation}\label{SchEffV}\begin{aligned}
        V_{\mathrm{Sch}}(r)&=\left(1-\frac{2 M}{r}\right)\left[\frac{\ell(\ell+1)}{r^2}+\frac{2 M\left(1-s^2\right)}{r^3}\right]
    \end{aligned}\end{equation}
    with the value of $s$ corresponding to the perturbation types:
    \begin{equation}
        s= \begin{cases}0, & \text {scalar perturbations}  \\
            1, & \text {vector perturbations}  \\
            2, & \text {tensor perturbations}
            \\\end{cases}
    \end{equation}
    
    Now, we can investigate the boundary behaviors of the eigen function by solving the equations as Eq.\eqref{eqBCs}, and the boundary condition for QNMs are:
    \begin{equation}
        R\rightarrow\begin{cases}
         \mathrm{e}^{-\mathrm{i}\omega r_*}& r_*\rightarrow-\infty(r\rightarrow 2M)  \\
            \mathrm{e}^{+\mathrm{i}\omega r_*} & r_*\rightarrow+\infty(r\rightarrow+\infty)  
            \\\end{cases}
    \end{equation}
    with $\mathrm{e}^{+\mathrm{i}\omega r_*}$ and $\mathrm{e}^{-\mathrm{i}\omega r_*}$ denoting outgoing and incoming waves respectively.
    
    There is a thorough discussion of the Schwarzschild spacetime in \cite{dewitt1973black} with two further reviews;~\cite{nollert1999quasinormal,kokkotas1999quasi}. Methods for calculating equations with QNMs are discussed in Sec.\ref{section:Method} and the reconstruction of the metric from the eigen function $R$ in Eq.\eqref{Scheqn} can be found in \cite{berti2009quasinormal}. The identical result can be obtained by using N-P formalism;~\cite{NP} (see details in \cite{chandrasekhar1975equations,chandrasekhar1984algebraically} and \cite{chandrasekhar1985mathematical} with the relationship between both of the multipoles for tensor perturbations can also be found).
    
    \subsection{Perturbations in Kerr Spacetime}\label{section:Kerr}
    
   Then, we discuss the perturbations of a static rotating axisymmetric BH as characterized generally by the Kerr solution;~\cite{kerr1963gravitational} in terms of the Boyer–Lindquist coordinate;~\cite{boyer1967maximal}:
    \begin{equation}
        \begin{aligned}
        \mathrm{d} s^2=&-\left(1-\frac{2 M r}{\Sigma}\right) \mathrm{d} t^2+\frac{\Sigma}{\Delta} \mathrm{d} r^2+\Sigma \mathrm{~d} \theta^2+\left(r^2+a^2+\frac{2 a^2 M r}{\Sigma} \sin ^2 \theta\right) \sin ^2 \theta \mathrm{d} \varphi^2\\& -\frac{4 a M r \sin ^2 \theta}{\Sigma} \mathrm{~d} t \mathrm{~d} \varphi
        \end{aligned}
    \end{equation}
    where $\Sigma \equiv r^2+a^2 \cos ^2\theta$, $\Delta \equiv r^2-2 M r+a^2$ and $a=\frac{J}{M}$ is the parameter describing the rotating property with $J$ the angular momentum.
    
    As with the difficulty of finding specific combinations of metric coefficients in Schwarzschild case, it is challenging to deal with relevant problems by using the metric perturbations methods. The metric in Kerr spacetime is determined by two parameters $M$ and $a$ in contrast with Schwarzschild case only $M$ remaining, which further complicates the problem. Separation of the dependence of $t$ and $\phi$ is obviously available due to the symmetry from stationarity and axisymmetry respectively;~\cite{teukolsky2015kerr}, meanwhile the discovery of the separability of $r$ and $\theta$ in scalar case brought hope for dealing with this problem;~\cite{carter1968hamilton}.
    
     However, it is also difficult to use metric perturbations methods to obtain a decoupled equation for Kerr case, fortunately the development of another method based on the N-P formalism;~\cite{NP} has proven its advantages for dealing with this problem in Schwarzschild case;~\cite{PhysRevD.5.2419,doi:10.1063/1.1666175}, thereby providing a superior method for analyzing such problems in Kerr case.
     
     In N-P formalism, one chooses four normalized orthogonal null vectors $\mathbf{l}, \mathbf{n}, \mathbf{m}, \mathbf{m}^*$ as the basis of a tetrad with the first two of those real and the remaining two being complex and conjugated with each other. The components of those in Boyer–Lindquist coordinate are given by;~\cite{teukolsky1973perturbations,chandrasekhar1985mathematical}:
    \begin{equation}
        \begin{aligned}
            l^a&=\frac{1}{\Delta}\left(r^{2}+a^{2},+\Delta, 0, a\right)\\ 
            n^a&=\frac{1}{2 \Sigma}\left(r^{2}+a^{2},-\Delta, 0, a\right)\\ 
            m^a&=\frac{1}{\sqrt{2}( r+i a \cos \theta)}(i a \sin \theta, 0,1, i \operatorname{cosec} \theta)\\
            (m^*)^a&=\frac{1}{\sqrt{2}( r-i a \cos \theta)}(-i a \sin \theta, 0,1, -i \operatorname{cosec} \theta) 
        \end{aligned}
    \end{equation}
     Following that, one may express the master equations and field quantities on N-P tetrad. For vector perturbations with the master equations given by Eq.\eqref{Maxwelleqn}, electromagnetic field tensor $F_{\mu\nu}$ can be donated in three independent complex scalar quantities $\phi_0,\phi_1,\phi_2$, where $\phi_0$ and $\phi_2$ describing the perturbations of a Maxwell field are defined as:
     \begin{equation}\label{NPVector}
        \begin{aligned}
        &\phi_{0}=F_{13}=F_{\mu \nu} l^{\mu} m^{\nu} \\
        &\phi_{2}=F_{42}=F_{\mu \nu} (m^*)^{\mu} n^{\nu}
        \end{aligned}
    \end{equation}
     while the vacuum Maxwell equations Eq.\eqref{Maxwelleqn} can be denoted in four equations.
     
    Similarly, the equations representing tensor perturbations in kerr spacetime Eq.\eqref{EinsteinEqn} become 18 N-P equations provided by Ricci identities and 8 complex equations derived from Bianchi identities. Meanwhile, Weyl tensors $C_{\mu\nu\sigma\lambda}$ turn to five N-P quantities $\Psi_0,\Psi_1,\Psi_2,\Psi_3,\Psi_4,$ and Ricci tensors are transformed into ten (including Ricci scalar), where $\Psi_0$ and $\Psi_4$ describing the incoming and outgoing gravitational radiation respectively are defined as:
    \begin{equation}\label{NPTensor}
        \begin{aligned}
            &\Psi_{0}=-C_{1313}=-C_{\mu\nu\sigma\lambda} l^{\mu} m^{\nu} l^{\sigma} m^{\lambda}\\
            &\Psi_{4}=-C_{2424}=-C_{\mu\nu\sigma\lambda} n^{\mu} (m^*)^{\nu} n^{\sigma} (m^*)^{\lambda}
        \end{aligned}
    \end{equation}
    
    
    By applying the separation of variables to the field with spin $s$:
    \begin{equation}\label{TeuFunction}
        \psi(t, r, \theta, \phi)=\frac{1}{2 \pi} \int \mathrm{e}^{-\mathrm{i} \omega t} \sum_{\ell=|s|}^{\infty} \sum_{m=-\ell}^\ell \mathrm{e}^{\mathrm{i} m \phi} {}_s S_{\ell m}(\theta) R_{\ell m}(r) d \omega
    \end{equation}
    with the specific form of the fields associated to different spin $s$ in Eq.\eqref{TeuFunction} called Teukolsky function shown in Table \ref{Table:TeuFunction}, one may obtain the decoupled equations for $r$ and $\theta$ respectively given by:
    \begin{equation}\label{rTeu}
        \left[\Delta^{-s} \frac{\mathrm{d}}{\mathrm{d} r}\left(\Delta^{s+1} \frac{\mathrm{d}}{\mathrm{~d} r}\right)+\frac{K^2-2 \mathrm{i} s(r-M) K}{\Delta}+4 \mathrm{i} s \omega r-{}_s\lambda_{\ell m}\right] R_{\ell m}=0
    \end{equation}
    and
    \begin{equation}\label{aTeu}
        \left[\frac{1}{\sin \theta} \frac{\mathrm{d}}{\mathrm{d} \theta}\left(\sin \theta \frac{\mathrm{d} }{\mathrm{~d} \theta}\right)+a^2 \omega^2 \cos ^2 \theta-2 a \omega s \cos \theta-\frac{(m+s \cos \theta)^2}{\sin ^2 \theta}+s+{}_sA_{\ell m}\right] {}_s S_{\ell m}=0
    \end{equation}
    with $K \equiv\left(r^2+a^2\right) \omega-a m$, ${}_s\lambda_{\ell m} \equiv {}_sA_{\ell m}+a^2 \omega^2-2 a m \omega$ and ${}_sA_{\ell m}$ the eigen value determined by the angular part equation Eq.\eqref{aTeu} produced from the separation of the dependence of $\theta$.
    
    The above decoupled equations are known as the \textbf{Teukolsky equations} and were first proposed and discussed by Teukolsky;~\cite{teukolsky1972rotating,teukolsky1973perturbations,press1973perturbations}, whose reasoning process can be seen in \cite{teukolsky2015kerr}, where he stressed the significance of a N-P variable $\tilde{\rho}$ written in Boyer–Lindquist coordinate as:
    \begin{equation}\label{NPrho}
        \tilde{\rho}=-\frac{1}{r-\mathrm{i} a \cos \theta}
    \end{equation}
    whose real and imaginary parts represent the divergence and curl of the outgoing principal null respectively. And the derivation process can be found from the Teukolsky's original essays as mentioned above, or from \cite{chandrasekhar1985mathematical}.

    The separation of the dependence of $\theta$ leads to the angular part equations Eq.\eqref{aTeu} with the eigen function ${}_s S_{\ell m}$ called \textbf{spin-weighted spheroidal harmonics (SWSH)} determined by the value of $s$, $\ell$, $m$ and $a\omega$. For $a \omega=0$ and $s=0$, it reduces to Schwarschild case with ${}_sA_{\ell m}=\ell(\ell+1)$ and ${}_sS_{\ell m}$ becoming the scalar spherical harmonics as defined in Eq.\eqref{aSch}. When $a \omega=0$ and $s\neq 0$, the eigen functions turn to spin-weighted spherical harmonics;~\cite{goldberg1967spin} with the eigen value ${}_sA_{\ell m}=\ell(\ell+1)-s(s+1)$. However, there is still no analytical solution for SWSH, therefore the determination of eigen values inevitably becomes a numerical problem; ~\cite{press1973perturbations,leaver1985analytic,seidel1989comment,berti2006eigenvalues} with the fully asymptotically behavior analysis can be found in \cite{hod2015eigenvalue}.
    
    The radial Teukolsky equations Eq.\eqref{rTeu} governing the QNMs in Kerr spacetime does not seem to have the same form as Eq.\eqref{QNMeqn}. However, under the transformation given by \cite{detweiler1977resonant} with the tensor case corrected in the Appendix B of \cite{maggio2021extreme}, those can be transformed into the form of Eq.\eqref{QNMeqn}. Along the same approach as in the Schwarzschild case, we can derive the following boundary asymptotic behavior;~\cite{teukolsky1974perturbations} at spatial infinity:
    \begin{equation}\label{TeuBCInf}
        R_{\ell m}\rightarrow
        \begin{cases}
          \frac{e^{-\mathrm{i} \omega r_*}} {r}& \text{incoming}  \\
            \frac{e^{+\mathrm{i} \omega r_*}}{r^{(2 s+1)}} & \text{outgoing}  \\
        \end{cases}\qquad r_*\rightarrow+\infty(r\rightarrow+\infty)
    \end{equation}
    and near horizon:
    \begin{equation}
        R_{\ell m}\rightarrow
        \begin{cases}
          \frac{\mathrm{e}^{-\mathrm{i} k r_*}}{\Delta^s} & \text{incoming}\\
            \mathrm{e}^{+\mathrm{i} k r_*} & \text{outgoing}  \\
        \end{cases} \qquad r_*\rightarrow-\infty(r\rightarrow r_+)
    \end{equation}
    with $k=\omega-m \omega_{+}$, $\omega_{+}=\frac{a} {2 M r_{+}}$ and $r_+$ the large root of $\Delta=0$ corresponds to the event horizon. In the case of QNMs of the BHs, the boundary condition should be chosen that $R_{\ell m}$ behaves incoming near horizon and outgoing at spatial infinity. Meanwhile, $r_*$ is the tortoise coordinate in Kerr spacetime defined as:
    \begin{equation}
        \frac{\mathrm{d} r_*}{\mathrm{d}r}=\frac{r^2+a^2}{\Delta}
    \end{equation}
    
    It is still challenging to reconstruct the metric coefficients, due to the difficulties to derive the QNMs equations in Kerr spacetime through another way, namely the metric perturbation method, who similarly prevents a direct relationship between the metric coefficients and the eigen functions of the QNMs equations. The only remaining option is to attempt to reconstruct the metric using Weyl tensors with $\Psi_0$ and $\Psi_4$ associated with the eigen functions of Teukolsky equations Eq.\eqref{TeuFunction} for $s=\pm 2$. However, because of the value of spin weight $s=\pm 2$, information on $\ell=0$ and $\ell=1$ associated with the perturbations of mass and angular momentum respectively are lost which must be provided by the rest of the N-P equations. Chandrasekhar attempted the construction and gave a set of methods in \cite{chandrasekhar1985mathematical}, but it was too complex for application in actual research.
    
    Another method called CCK procedure is based on a key result first proposed by \cite{chrzanowski1975vector} and developed by \cite{PhysRevLett.41.203,stewart1979hertz,PhysRevD.19.1641} where they reconstructed the metric perturbation $h_{\mu\nu}$ from a spin-2 scalar \emph{Hertz potential} with the adoption of radiation gauge;~\cite{PhysRevD.64.124003}. The first case of metric reconstruction for the nonvacuum situation is given by \cite{PhysRevD.67.124010} with some other applications can be found in \cite{yunes2006metric,sano2014gravitational,merlin2016completion} and a relatively thorough overview of the procedure can be found in \cite{van2017mass,toomani2021new}. Furthermore, a recent research seeks to expand the method to the general Lorentzian gauge in order to address the singularity problem in nonvacuum situations;~\cite{dolan2022gravitational}. Meanwhile, \cite{loutrel2021second} attempt to develop a new method to reconstruct the first-order metric perturbation just from the solution of the first-order Teukolsky equation, without the requirement for Hertz potentials.
    
    It is worth mentioned that the solutions of the homogeneous QNMs equations will exponentially diverge near infinity for example in Schwarzschild and Kerr cases as shown in Fig.\ref{fig:SchBC}, which brings more difficulties in calculation. Fortunately, another equivalent form to Teukolsky equation Eq.\eqref{rTeu} which is more friendly to numeral calculation was developed in \cite{sasaki1982gravitational}.
    
    \subsection{Other Cases}
    
    We summarize the publications where the formalism of the QNMs governing equations for gravitational perturbation can be found.
    
    \begin{itemize}
        \item The QNMs governing equation of the axial gravitational perturbation in general spherical symmetric spacetime can be found in \cite{zhang2021gravitational}, while those of polar parity can be found in \cite{liu2022gauge}.
        
        \item The perturbation equations for tensor case was discussed in Kerr-Newman-de Sitter spacetime ;~\cite{suzuki1998perturbations}.
        
        \item  Extension to the anti-de Sitter spacetime for Schwarzschild case can be found in \cite{cardoso2001quasinormal}, while those of Kerr can be found in \cite{tattersall2018kerr} under the slow rotation limit.
        
        \item Extension to higher dimensions cases can be found in \cite{kodama2003master,ishibashi2003stability,kodama2004master}.
        
        \item The discussion for the linear tensor perturbations under several modified gravity theories can be found in \cite{moulin2019overview}.
    \end{itemize}

\section{Methods for Calculating QNMs}\label{section:Method}

    In the previous section, we reviewed linear perturbation theory and the related QNMs equations. We have turned the problem of the perturbations in the curved spacetime into a set of Schrodinger-like equations Eq.\eqref{QNMeqn} with the corresponding boundary conditions, specifically incoming at the horizon and outgoing at spatial infinity. In this section, we will discuss how to solve these equations and obtain the accurate eigenvalues or QNMs.
    
    This seems to be a straightforward eigen value problem: directly integrate from one boundary to the other and use the shooting method (like \cite{chandrasekhar1975quasi}) to obtain the appropriate eigen values. However, when one does so, the exponentially diverging asymptotic behaviors on the boundaries cause the numerical error to increase exponentially, which contradicts the requirement of the common shooting method to increase the value of $r_*$ (or $r$) as large as possible to match the asymptotic behavior at infinity, making it difficult to find accurate QNMs directly using numerical integration. In other words, singularities at the horizon and spatial infinity bring the difficulties of direct integration significantly by using shooting method. Fig.\ref{fig:SchBCH} and Fig.\ref{fig:SchBCInf} depict the boundary behaviors of Schwarzschild case for example to illustrate the asymptotic behaviors of exponential growth at the boundary.
    
    
    The analysis of the challenge of numerically of locating accurate QNMs producing from asymptotic boundary behavior can be found in \cite{nollert1992quasinormal} who also introduced Green function methods to deal with the inhomogeneous equations resulting in a series of studies in nonvacuum case, with a thorough discussion can be found in \cite{poisson2011motion}. 
    
     In fact, the employment of analytical or semi-analytical methods to supplement purely numerical methods may simplify and improve the processing of related problems. We will illustrate the general idea of solving this problem by discussing two well known methods: WKB approximation methods and continue fraction methods. Meanwhile, some other methods are listed at the end of this section.
    
    \subsection{WKB Approximation Methods}
    
    WKB (also known as JWKB) approximation methods were fist proposed by \cite{jeffreys1925certain} with a general method of employing approximate solutions to solve linear second order differential equations including the Schrödinger equation, and developed by \cite{wentzel1926verallgemeinerung,kramers1926wellenmechanik,brillouin1926mecanique} with the treatment of the turning points to address specific problems in quantum mechanics. Meanwhile, the fundamental concepts of the WKB method are usually summarized in almost every quantum mechanics literature;~\cite{froman1965jwkb,hall2013quantum}.
    
    Since its first application in the perturbations problem of a BH;~\cite{schutz1985black}, this method has undergone constant development, and it remains one of the most effective methods for exploring related problems. For convenience, we rewrite the QNMs governing equation Eq.\eqref{QNMeqn} in another form:
    \begin{equation}\label{WKB:QNMeqn}
        \left(\epsilon^2\frac{\mathrm{d}^2}{\mathrm{d} x^2}+Q(x)\right) \Psi(x)= 0
    \end{equation}
    where $Q(x)=\omega^2-V_{\text{eff}}$ and $\epsilon$ is a small parameter to track the order of WKB approximation first introduced by \cite{iyer1987black} during his research for third order WKB method. After setting $\epsilon=1$, Eq.\eqref{WKB:QNMeqn} returns to its original form Eq.\eqref{QNMeqn}.
    
    
    The WKB approximation retains high precision only in the so-called classically allowed region defined by $Q(x)>0$. Considering that $Q(x)$ (or $V_{\text{eff}}$) is usually unimodal, $Q(x)\sim 0$ produces two turning points and divides the whole integration domain into three regions as shown in Fig.\ref{fig:WKB}. In regions $\mathrm{I}$ and $\mathrm{III}$, the WKB approximation is introduced by assuming the solution in the form of the asymptotic series expansion of $\epsilon$ as:
    \begin{equation}\label{WKB:ansatz}
        \Psi \sim \exp \left[\sum_{n=0}^{\infty} \frac{S_n(x) \epsilon^n}{\epsilon}\right]
    \end{equation}
    by substituting the ansatz Eq.\eqref{WKB:ansatz} into Eq.\eqref{WKB:QNMeqn} and equating the same powers of $\epsilon$, the specific form of $S_n$ can be solved order by order. For example, the fundamental and the first order solutions can be solved as:
    \begin{equation}\label{WKB:S0}
        S_0(x)=\pm \mathrm{i} \int^x \sqrt{Q(\eta)} \mathrm{d} \eta
    \end{equation}
    and
    \begin{equation}
        S_1(x)=-\frac{1}{4} \ln Q(x)
    \end{equation}
    with the sign of Eq.\eqref{WKB:S0} determined by the asymptotic behavior taken at both of the boundaries. Under the consideration of only the fundamental solution with $\Psi\sim \mathrm{e}^{S_0}$, the boundary behavior of $\Psi\sim\mathrm{e}^{\pm\mathrm{i}\omega x}$ corresponds to $S_0\sim \pm\mathrm{i}\omega x$. Thus, after introducing the four solutions $\Psi_+^\mathrm{I}$, $\Psi_-^\mathrm{I}$, $\Psi_+^\mathrm{III}$ and $\Psi_-^\mathrm{III}$ to denote the corresponding signs in regions $\mathrm{I}$ and $\mathrm{III}$ respectively with the boundary behaviors in region $\mathrm{I}$ (spatial infinity) of Fig.\ref{fig:WKB} as:
   \begin{equation}
        \begin{cases}
          \Psi_-^\mathrm{I}\sim\mathrm{e}^{-\mathrm{i}\omega x}& \text{in}  \\
            \Psi_+^\mathrm{I}\sim\mathrm{e}^{+\mathrm{i}\omega x}& \text{out}  \\
        \end{cases}\qquad \text{region}\;\mathrm{I}\;(x\rightarrow+\infty)
    \end{equation}
    and in region $\mathrm{III}$ (horizon):
    \begin{equation}
        \begin{cases}
          \Psi_-^\mathrm{III}\sim\mathrm{e}^{-\mathrm{i}\omega x}& \text{out}  \\
            \Psi_+^\mathrm{III}\sim\mathrm{e}^{+\mathrm{i}\omega x}& \text{in}  \\
        \end{cases}\qquad \text{region }\;\mathrm{III}\;(x\rightarrow-\infty)
    \end{equation}
    where the above "in" and "out" represent the waves incident from region $\mathrm{I}$ (or region $\mathrm{III}$) to region $\mathrm{II}$ and the waves emitted from region $\mathrm{II}$ to region $\mathrm{I}$ (or region $\mathrm{III}$) respectively (not the incoming and outgoing waves). We then obtain the general solutions in regions $\mathrm{I}$ and $\mathrm{III}$ given as:
    \begin{equation}\label{WKB:sol13}
    \Psi\sim
        \begin{cases}
          Z_{\text{in}}^\mathrm{I} \Psi_{-}^\mathrm{I}+Z_{\text{out}}^\mathrm{I} \Psi_{+}^\mathrm{I}& \text{region}\;\mathrm{I}  \\
            Z_{\text{in}}^\mathrm{III} \Psi_{+}^\mathrm{III}+Z_{\text{out}}^\mathrm{III} \Psi_{-}^\mathrm{III}& \text{region}\;\mathrm{III}  \\
        \end{cases}
    \end{equation}
    And the amplitudes in region $\mathrm{I}$ are associated with those in region $\mathrm{III}$ through the linear matrix:
    \begin{equation}\label{WKB:matrix}
        \left(\begin{array}{c}
        Z_{\text {out }}^{\mathrm{III}} \\
        Z_{\text{in}}^{\mathrm{III}}
        \end{array}\right) \equiv\left(\begin{array}{ll}
        M_{11} & M_{12} \\
        M_{21} & M_{22}
        \end{array}\right)\left(\begin{array}{c}
        Z_{\text {out }}^{\mathrm{I}} \\
        Z_{\text {in }}^{\mathrm{I}}
        \end{array}\right)
    \end{equation}
    where $M_{11}$, $M_{12}$, $M_{21}$ and $M_{22}$ are the coefficients determined by the matching of WKB solutions Eq.\eqref{WKB:sol13} in regions $\mathrm{I}$ and $\mathrm{III}$ with the solution in region $\mathrm{II}$ respectively.
    
    The determination of the elements of the matrix in Eq.\eqref{WKB:matrix} needs to consider the solution in region $\mathrm{II}$ by approximating $Q(x)$ in the form of Taylor series at the peak of $Q(x)$ as:
    \begin{equation}
        Q(x)=Q_0+\frac{1}{2}Q_0^{\prime\prime}\left(x-x_0\right)^2+O\left(\left(x-x_0\right)^3\right)
    \end{equation}
    with $x_0$ the point of the maximum of $Q(x)$, $Q_0=Q(x_0)$ and $Q_0^{\prime\prime}$ the second derivative with respect to $x$ at the point $x=x_0$. And the above Taylor expansion approximation is valid under the assumption that $\left|x-x_0\right|$ is a small value, or to be exact considering the scope of region $\mathrm{II}$ given by:
    \begin{equation}\label{WKB:region2}
        \left|x-x_0\right|<\sqrt{\frac{-2 Q_0}{Q_0^{\prime \prime}}} \approx \sqrt{\epsilon}
    \end{equation}
    with $\epsilon$ a small value which also gives the validity of the approximation. After that Eq.\eqref{WKB:QNMeqn} can be rewritten in the form of parabolic cylinder equation;~\cite{bender1978sa,olver2010nist}:
    \begin{equation}
        \left(\frac{\mathrm{d}^2}{\mathrm{d} t^2}+\nu+\frac{1}{2}-\frac{1}{4} t^2\right) \Psi=0
    \end{equation}
    with the substitution as;~\cite{iyer1987black}:
    \begin{equation}
        k = \frac{1}{2} Q_0^{\prime \prime},\qquad
        t=(4 k)^{\frac{1}{4}}\mathrm{e}^{\frac{-\mathrm{i}\pi}{4}}\left(x-x_0\right)\frac{1}{\sqrt{\epsilon}} 
    \end{equation}
    \begin{equation}\label{WKB:n}
        z_0^2 =\frac{-2 Q_0}{Q_0^{\prime \prime}},\qquad
        \nu+\frac{1}{2} =\frac{-\mathrm{i} \sqrt{k} z_0^2}{2}\frac{1}{\epsilon}
    \end{equation}
    And the general solution of this equation can be denoted as the linear combination of the parabolic cylinder functions as:
    \begin{equation}\label{WKB:sol2}
        \Psi=A D_\nu(t)+B D_{-\nu-1}(i t)
    \end{equation}
    Under the asymptotic behavior of the parabolic cylinder functions, the solutions become:
    \begin{equation}\label{WKB:2inf}
        \begin{aligned}
        \Psi \sim& B \mathrm{e}^{\frac{-3 \mathrm{i} \pi(\nu+1)}{4}}(4 k)^{-\frac{\nu+1}{4}}\left(x-x_0\right)^{-(\nu+1)} \mathrm{e}^{\frac{\mathrm{i} \sqrt{k}\left(x-x_0\right)^2}{2}} \\
        &+\left(A+B\frac{(2 \pi)^{1 / 2 \mathrm{e}^{-i \nu \pi / 2}}}{\Gamma(\nu+1)}\right)  \\
        &\times \mathrm{e}^{\frac{\mathrm{i} \pi \nu}{4}}(4 k)^{\frac{\nu}{4}}\left(x-x_0\right)^\nu \mathrm{e}^{\frac{-\mathrm{i} \sqrt{k}\left(x-x_0\right)^2}{2}}, \quad x \gg x_2,
        \end{aligned}
    \end{equation}
    and
    \begin{equation}\label{WKB:2horizon}
        \begin{aligned}
        \Psi \sim& A \mathrm{e}^{-3 \mathrm{i} \pi \nu / 4}(4 k)^{\nu / 4}\left(x-x_0\right)^\nu \mathrm{e}^{-\mathrm{i} \sqrt{k}\left(x-x_0\right)^2 / 2}\\
        &+\left(\frac{B-\mathrm{i} A(2 \pi)^{1 / 2} \mathrm{e}^{-\mathrm{i} \nu \pi / 2}}{\Gamma(-\nu)}\right) \mathrm{e}^{\frac{\mathrm{i} \pi(\nu+1)}{4}}(4 k)^{\frac{-(\nu+1)}{4}} \\
        &\times\left(x-x_0\right)^{-(\nu+1)} \mathrm{e}^{\mathrm{i} \sqrt{k}\left(x-x_0\right)^2 / 2}, \quad x \ll x_1
        \end{aligned}
    \end{equation}
    where $x_1$ is the smaller turning point and $x_2$ is the bigger one as shown in Fig.\ref{fig:WKB}.
    
    Now that we have calculated the asymptotic solutions close to both of the turning points on both sides, we must match them. Specifically, around the bigger turning point that serves as the dividing point between regions $\mathrm{I}$ and $\mathrm{II}$, we must match the coefficients of Eq.\eqref{WKB:2inf} with those of Eq.\eqref{WKB:sol13} in region $\mathrm{I}$, while we do the same thing near another turning point by matching the coefficients of Eq.\eqref{WKB:2horizon} with those of Eq.\eqref{WKB:sol13} in region $\mathrm{III}$. And after eliminating the coefficients $A$ and $B$, we obtain the elements of the matrix in Eq.\eqref{WKB:matrix} only based on $\nu$:
    \begin{equation}\label{WKB:matrix1}
        \left(\begin{array}{c}
        Z_{\text{out}}^{\mathrm{III}} \\
        Z_{\text{in}}^{\mathrm{III}}
        \end{array}\right)=\left(\begin{array}{cc}
        \mathrm{e}^{\mathrm{i} \pi \nu} & \frac{\mathrm{i} R^2 \mathrm{e}^{\mathrm{i} \pi \nu}(2 \pi)^{1 / 2}}{\Gamma(\nu+1)} \\
        \frac{R^{-2}(2 \pi)^{1 / 2}}{\Gamma(-\nu)} & -\mathrm{e}^{\mathrm{i} \pi \nu}
        \end{array}\right)\left(\begin{array}{c}
        Z_{\text{out}}^{\mathrm{I}} \\
        Z_{\text{in}}^{\mathrm{I}}
        \end{array}\right)
    \end{equation}
    with
    \begin{equation}\label{WKB:R}
        R=(\nu+\frac{1}{2})^{\frac{1}{2}(\nu+\frac{1}{2})} \mathrm{e}^{-\frac{1}{2}(\nu+\frac{1}{2})}
    \end{equation}
    For QNMs case of a BH, the condition of the normal modes limits the coefficients in region $\mathrm{I}$ with $Z_{\text{in}}^{\mathrm{I}}=0$, while the BHs indicate that there is no wave reflected from the horizon with $Z_{\text{in}}^{\mathrm{III}}=0$. By applying the above conditions to Eq.\eqref{WKB:matrix1}, we obtain the limitation:
    \begin{equation}
        \frac{1}{\Gamma(-\nu)}=0
    \end{equation}    
    and $\nu$ must be an integer corresponding to the overtone number $n$. After considering Eq.\eqref{WKB:n}, we obtain the QNMs under the first order of WKB approximation determined by:
    \begin{equation}\label{WKB:QNMs}
        n+\frac{1}{2}=\frac{\mathrm{i} (\omega^2-V_{0})}{\sqrt{2 Q_0^{\prime \prime}}}, \quad n=0,\pm 1,\pm 2, \ldots
    \end{equation}
    where $V_0$ is the peak of the effective potential $V_{\text{eff}}$ and the signs of $n$ denote the real part of $\omega$ as:
    \begin{equation}
        n= \begin{cases}0,1,2, \ldots, & \operatorname{Re} \omega>0 \\ -1,-2, \ldots, & \operatorname{Re} \omega<0\end{cases}
    \end{equation}
    
    \textbf{For a concise summary}, to explore such problems using the WKB methods, the following stages are usually taken:
    
    \begin{itemize}
    
        \item The whole integration domain is divided into several regions by the turning points determined by $Q(x)\sim 0$, as the unimodal potential shown in Fig.\ref{fig:WKB}.
        
        \item By applying WKB approximation Eq.\eqref{WKB:ansatz} to the QNMs governing equations Eq.\eqref{WKB:QNMeqn} in the regions $\mathrm{I}$ and $\mathrm{III}$ determined by $Q(x)>0$, we then obtain
        general solution in these regions as Eq.\eqref{WKB:sol13}.
                
        \item By approximating $Q(x)$ in region $\mathrm{II}$ using Taylor expansion and rewriting the equation into analytical parabolic cylinder equation, we obtain the general solutions in the form of the linear combination of parabolic cylinder functions Eq.\eqref{WKB:sol2} with the asymptotic behaviors near the turning points as Eq.\eqref{WKB:2inf} and Eq.\eqref{WKB:2horizon}.
        
        \item By matching the corresponding coefficients and eliminating the $A$ and $B$ near different turning points, we obtain the matrix in Eq.\eqref{WKB:matrix1}.
        
        \item By considering the substitution Eq.\eqref{WKB:n} and applying the corresponding coefficients according to the specific physical problem such as $Z_{\text{in}}^{\mathrm{I}}=Z_{\text{in}}^{\mathrm{III}}=0$, we then obtain the first WKB order estimate values of QNMs determined by Eq.\eqref{WKB:QNMs}.
        
    \end{itemize}
    
    The higher WKB approximation methods lead to the same form of Eq.\eqref{WKB:matrix1}, with the modified expression for $R$ in Eq.\eqref{WKB:R} still only based on $\nu$. In the meanwhile, the higher order Taylor expansion series result in the different substitution of $\nu$ in Eq.\eqref{WKB:n} that leads to QNMs determined by:
    \begin{equation}
        n+\frac{1}{2}=\frac{\mathrm{i} (\omega^2-V_{0})}{\sqrt{2 Q_0^{\prime \prime}}}-\sum_{i=2}\Lambda_i, \quad n=0,\pm 1,\pm 2, \ldots
    \end{equation}
    where $\Lambda_i$ are the functions of the values of the effective potential and the derivatives (up to the $i\text{-th}$ order) at the maximum of the effective potential. The explicit modified terms $\Lambda_2$, $\Lambda_3$ of third WKB order approximation methods can be found in \cite{iyer1987black} with the calculation of QNMs;~\cite{iyer1987black2,kokkotas1988black,seidel1990black}. Meanwhile, $\Lambda_4$, $\Lambda_5$ and $\Lambda_6$ for sixth WKB order approximation can be found in \cite{konoplya2003quasinormal,konoplya2004quasinormal}. And those of the thirteenth WKB order approximation were 
   provided by \cite{matyjasek2017quasinormal}, with introducing Padé approximation instead of the Taylor series, leading to more accurate results than those of the sixth WKB order in several cases;~\cite{konoplya2019higher}.

    \subsection{Continued Fraction Methods (Leaver's Methods)}
    
    The study of continued fraction in mathematics goes back hundreds of years, however, it was not introduced in the eigen value problems until 1934 by \cite{jaffe1934theorie} where the bound state of the hydrogen molecule ion;~\cite{hylleraas1931elektronenterme} was studied and he obtained a solution with the proof of convergence, while the same discovery was made by \cite{baber1935two} independently. Meanwhile, the early related works were reviewed in \cite{leaver1986solutions}.
    
    With the observation;~\cite{leaver1986solutions} that the Teukolsky equations are the subclass of spheroidal wave equations arising during the process of \cite{jaffe1934theorie,baber1935two}, Leaver first introduced the continue fraction into the linear perturbation problems and calculated the QNMs in Schwarzschild and Kerr spacetime in \cite{leaver1985analytic}. After decades of development, this method is one of the most effective ways for estimating QNMs and can provide almost the most accurate value of QNMs. We will illustrate the general thought in Kerr spacetime with the unit $c=G=2M=1$ the same as \cite{leaver1985analytic}:
    
    Instead of transforming the equations into the form of Eq.\eqref{QNMeqn} in the tortoise coordinate, we often discuss the Teukolsky equations Eq.\eqref{rTeu} directly in $r$ coordinates with the boundary conditions for QNMs of a Kerr BH given as:
    \begin{equation}
        R_{\ell m}\sim
        \begin{cases}
         \left(r-r_{+}\right)^{-s-\mathrm{i} \sigma_{+}}& r\rightarrow r_+\\
            r^{-1-2 s+\mathrm{i} \omega} \mathrm{e}^{\mathrm{i} \omega r} &r\rightarrow +\infty 
        \end{cases}
    \end{equation}
    with $\sigma_{+}=\frac{\omega r_{+}-a m}{\sqrt{1-4 a^2}}$. Following the Leaver's approach, we assume the expression of the Teukolsky functions being finite at the regular singular points or the boundaries as:
    \begin{equation}
        R_{\ell m}=\mathrm{e}^{\mathrm{i} \omega r}\left(r-r_{-}\right)^{-1-s+\mathrm{i} \omega+\mathrm{i} \sigma_{+}}\left(r-r_{+}\right)^{-s-\mathrm{i} \sigma_{+}} \sum_{k=0}^{\infty} a_k^{r}\left(\frac{r-r_{+}}{r-r_{-}}\right)^k
    \end{equation}
    By substituting the above ansatz into the radial Teukolsky equations Eq.\eqref{rTeu} and equating the coefficients of each orders to zero, the expression coefficients satisfy the following three-term recursion relation:
    \begin{equation}\label{Leaver:0BC}
        \alpha_0^r a_1^{r}+\beta_0^r a_0^{r}=0
    \end{equation}
    and
    \begin{equation}\label{Leaver:3term}
        \alpha_k^r a_{k+1}^{r}+\beta_k^r a_k^{r}+\gamma_k^r a_{k-1}^{r}=0, \quad k=1,2 \ldots
    \end{equation}
    where $\alpha_k^r$, $\beta_k^r$ and $\gamma_k^r$ are the recursion coefficients functions of $k$, $\omega$ and $a$, $s$, $m$, $A_{\ell m}$ as for the parameters of the QNMs governing equations Eq.\eqref{rTeu}, with the specific formalism can be found in \cite{leaver1985analytic}. Then, it turns to the problem of dealing with the three-term recursion relation whose properties explored by \cite{gautschi1967computational}. Eq.\eqref{Leaver:3term} leads to the continued fraction which determines the values of QNMs $\omega$ as:
    \begin{equation}
        R_k=-\frac{a_{k+1}^r}{a_k^r}=\frac{\gamma_{k+1}}{\beta_{k+1}-} \frac{\alpha_{k+1} \gamma_{k+2}}{\beta_{k+2}-} \frac{\alpha_{k+2} \gamma_{k+3}}{\beta_{k+3}-} \ldots
    \end{equation}
    where the continued fraction $R_k$ can be regarded as the function of $\omega$ for given $a$, $s$, $m$ and $A_{\ell m}$ with the boundary conditions as $k\rightarrow \infty$ and $k=0$. The analysis of the convergence of the expansion coefficients as $k\rightarrow \infty$ indicates;~\cite{gautschi1967computational}:
    \begin{equation}\label{Leaver:BCinf}
        -R_k=\frac{a_{k+1}^r}{a_k^r} \rightarrow 1 \pm \sqrt{\frac{-2 \mathrm{i}\omega}{k}}-\frac{8 \mathrm{i}\omega+3}{4k}+\ldots\rightarrow 1,\qquad k\rightarrow \infty
    \end{equation}
    Meanwhile, boundary condition at $k=0$ are given from Eq.\eqref{Leaver:0BC}. By substituting it into the continued fraction for $R_0$, we obtained the characteristic equation determining the QNMs as:
    \begin{equation}\label{Leaver:QNMs1}
        0=\beta_0-\frac{\alpha_0 \gamma_1}{\beta_1-} \frac{\alpha_1 \gamma_2}{\beta_2-} \frac{\alpha_2 \gamma_3}{\beta_3-} \ldots
    \end{equation}
    with the equivalent formalism by inverting an arbitrary number of times $k$ given as:
    \begin{equation}\label{Leaver:QNMs2}
        \beta_k-\frac{\alpha_{k-1} \gamma_k}{\beta_{k-1}-} \frac{\alpha_{k-2} \gamma_{k-1}}{\beta_{k-2}-} \ldots-\frac{\alpha_0 \gamma_1}{\beta_0}=\frac{\alpha_k \gamma_{k+1}}{\beta_{k+1}-} \frac{\alpha_{k+1} \gamma_{k+2}}{\beta_{k+2}-} \frac{\alpha_{k+2} \gamma_{k+3}}{\beta_{k+3}-} \ldots \quad(k=1,2 \ldots)
    \end{equation}
    For given $a$, $s$, $m$, $A_{\ell m}$, and by setting $k=k_c$ in a large cutoff value with $R_{k_c}=1$ due to the boundary condition from Eq.\eqref{Leaver:BCinf}, QNMs $\omega$ now become the roots of Eq.\eqref{Leaver:QNMs1} or Eq.\eqref{Leaver:QNMs2} and can be obtained through a numeral method.
    
    However, the determination of $A_{\ell m}$ may be a nontrivial problem as is mentioned in Sec.\ref{section:Kerr}. Following the same approach, the angular Teukolsky equations Eq.\eqref{aTeu} can be solved by supposing the series solution for the angular eigen functions as:
    \begin{equation}
        S_{\ell m}(u)=\mathrm{e}^{a \omega u}(1+u)^{\frac{1}{2}|m-s|}(1-u)^{\frac{1}{2}|m+s|} \sum_{k=0}^{\infty} a_k^{\theta}(1+u)^k
    \end{equation}
    with $u=\cos\theta$. And it turns to the similar three-term recursion relation as:
    \begin{equation}
        \alpha_0^\theta a_1^{\theta}+\beta_0^\theta a_0^{\theta}=0
    \end{equation}
    \begin{equation}
        \alpha_k^\theta a_{k+1}^{\theta}+\beta_k^\theta a_k^{\theta}+\gamma_k^\theta a_{k-1}^{\theta}=0, \quad k=1,2 \ldots
    \end{equation}
    where the form of the corresponding coefficients $\alpha_k^\theta$, $\beta_k^\theta$ and $\gamma_k^\theta$ can be found in \cite{leaver1985analytic}. And the eigen value $A_{\ell m}$ for given $a$, $s$ and $m$ can be obtained from the above method.
    
   As the overtone value $n$ increases, the convergence of continued fraction worsens as well \cite{starinets2002quasinormal} which leads to the calculation for higher overtone requiring larger cutoff value $k_c$ with more computing power. Based on this difficulty, a method applicable to higher overtone QNMs was generalized by expand continued fraction $R_k$ in the series of $\frac{1}{\sqrt{k}}$;~\cite{nollert1993quasinormal,zhidenko2006massive}.
   
   Meanwhile, the continued fraction from the Frobenius series can be found in \cite{konoplya2011quasinormal}. And the application of this method in several cases can be found in \cite{leaver1985analytic,onozawa1997detailed,berti2004highly} for Kerr BHs, in \cite{leaver1990quasinormal} for Reissner-Nordström BHs, in \cite{berti2005quasinormal} for Kerr-Newman BHs.
   
    \subsection{Other Methods}
    
    Due to the complexity of the gravity theories and the resulting spacetime geometries, in many cases, we must deal with the corresponding perturbation problems case by case. We summarize some publications that employ additional methods and asymptotic formalism.
    
    \begin{itemize}
    
     \item \textbf{Shooting Method and Its Extension}. For this kind of eigenvalue problem, the obvious option is to integrate directly and use the shooting method. The most straightforward strategy is to integrate directly from horizon to a large cutoff value with equating the coefficient of the outgoing wave to zero;~\cite{press1973perturbations}.
        
            The effective extension of this method can be found in \cite{chandrasekhar1975quasi}, where he applied Taylor series expansion at the horizon and spatial infinity respectively, before integrating to a specific intermediate point and matching the both solutions by equating the Wronskian of them to zero, while the details and extension in a matrix formalism can be found in \cite{molina2010gravitational}. An interesting example using this method can be found in \cite{mai2022extremal} where the unstable QNMs in a special gravity theory were found.
    
    \item \textbf{The ‘Phase–amplitude’ Method} This method try to deal directly with singularities that appear with $r\rightarrow+\infty$ by choosing a specific integral curve in the complex plane to make numerical integration methods possible;~\cite{froman1992black}, while \cite{andersson1992numerically} use this method to calculate QNMs in Schwarzschild case.
    
    \item \textbf{Exact Solutions for Special Potentials} Even though it is challenging to obtain exact solutions for the actual BHs QNMs equations, there are always Schrodinger-like equations with special potentials associated to analytic solutions. The researches in \cite{BLOME1984231} and \cite{PhysRevLett.52.1361} analyzed the connection between QNMs and bound states of the inverted effective potential while used inverted \emph{Pöschl-Teller potential} to approximate the actual BHs effective potentials in Schwarzschild, Kerr and Reissner-Nordström cases. Some of the other useful potentials can be found in \cite{boonserm2011quasi}.
        
    \item \textbf{Exact Solutions of Heun Equation} The Heun equation is a generalization of the hypergeometric equation while the radial and angular perturbation equations can be rewritten in the form of confluent Heun equations ;~\cite{arscott1995heun,baumann2019spectra}, with related researches in \cite{fiziev2006exact} for Schwarzschild case and in \cite{borissov2009exact,fiziev2009teukolsky} for Kerr case. However, accurate and fast calculation of the Heun functions are also the limitation of this method, fortunately the implement of the Heun functions in the \emph{Mathematica} 12.1 make it possible to obtain a precise solution in a few seconds;~\cite{hatsuda2020quasinormal}.
         
    \item \textbf{Post-Newtonian (PN) Expansion Method.} This method based on post-Newtonian expansion is extensively employed in the calculation of the GWs waveform for a binary system;~\cite{GWReview1,futamase2007post,cho2022third}, thus the applicability to perturbation problems seems evident. The details of this method can be found in \cite{10.1143/PTPS.128.1,sasaki2003analytic}. The examples for Schwarzschild case can be found in \cite{fujita2012gravitational} with 22PN expansion and for Kerr in \cite{fujita2015gravitational} with 11PN expansion. However, such accuracy is still not enough for GWs detection;~\cite{sago2016accuracy}.
        
    \item \textbf{Asymptotic Iteration Method (AIM)} This method is based on a mathematical theorem resulting in a equivalently condition for the linear homogeneous second order ordinary differential equations, including the QNMs equations. By dealing with the above equivalently condition in numeral methods, the accurate QNMs can be found. The details are reviewed in \cite{ciftci2003asymptotic,cho2012new} with the calculation of QNMs can be found in \cite{mamani2022revisiting} for Schwarzschild and the available \emph{Julia} package in \cite{sanches2022quasinormalmodes}. 
    
    \item \textbf{The Pseudo-spectral Method} In this method, the continuous independent variables (radial coordinate for QNMs equations) are replaced by a discrete set of points called the grid, thus the eigen function can be approximated by a series of cardinal functions corresponding to the grid;\cite{jansen2017overdamped}. Then, the coefficients of each order of eigen functions can be expanded as the series of $\omega$ which results in a matrix governing the eigenvalue problem. The calculation of QNMs can be found in \cite{mamani2022revisiting} for Schwarzschild and the details with available \emph{Mathematica} package in \cite{jansen2017overdamped},
    
    \item \textbf{MST Method} This method is based on the formalism developed by Mano, Suzuki and Takasugi;~\cite{10.1143/PTP.95.1079}. The homogeneous radial Teukolsky solutions in Eq.\eqref{rTeu} are expanded in the series of hypergeometric functions near horizon and Coulomb wave functions at spatial infinity, resulting in the three term recurrence relation for their expansion coefficients respectively, which is similar to Leaver's method;~\cite{leaver1986solutions}. However, the both three term recurrence relations among the expansion coefficients are the same which makes the analytically match possible;~\cite{10.1143/PTP.112.415}. The details can be found in \cite{10.1143/PTP.112.415,10.1143/PTP.121.843} with the program implemented in \cite{BHPToolkit} and an example can be found in \cite{PhysRevD.102.024041}.
    
    \end{itemize}
    
\section{Black Hole Spectroscopy and Detection Advancements}\label{Section:detection}

    In the previous sections, we reviewed linear perturbation theory leading to the related QNMs equations in Sec.\ref{section:QNMeqn} and the calculation of QNMs in Sce.\ref{section:Method}. Then, we wonder what the perturbed BHs at a great distance may appear like on our detectors, accurately speaking, they are the GWs waveform during the ringdown stage, which leads to the detection of BHs spectroscopy;~\cite{berti2006gravitational,berti2007matched,berti2005estimating}. 
    
    One of the most common states of a perturbed BH is the remnant of the final ringdown stage of a binary BH merger event, which can be regarded as the perturbation of a rotating black problem described by the Teukolsky equations Eq.\eqref{aTeu} and Eq.\eqref{rTeu}. Meanwhile the charges of a BH have been proven to have no detectable effect on the ringdown waveform;~\cite{PhysRevD.105.062009}. Due to the difficulties of reconstructing the metric in terms of Teukolsky functions as is mentioned at the end of Sec.\ref{section:Kerr}, it is difficult to obtain the GWs waveform by directly applying TT gauge;~\cite{GWreview2}. However, the asymptotic behavior of $\Psi_4$ at infinity is naturally associated with the both polarization modes of the outgoing GWs;~\cite{teukolsky1973perturbations}:
    \begin{equation}
        \Psi_4=-\frac{1}{2}(\ddot h_+ + \mathrm{i} \ddot h_\times)=-\frac{1}{2}\omega^2(h_+ + \mathrm{i} h_\times), \qquad r\sim +\infty
    \end{equation}
    where $h_+=h_{\theta\theta}$ and $h_\times=h_{\theta\phi}$ are the both polarization modes and the dots on the top denote the derivative with respect to time $t$. With the relation between $\Psi_4$ and the Teukolsky functions Eq.\eqref{TeuFunction} provided in Table \ref{Table:TeuFunction}, the GWs waveform at infinity can be written in;~\cite{berti2006gravitational}:
    \begin{equation}
        h_{+}+\mathrm{i} h_{\times}=-\frac{2}{r^4} \int_{-\infty}^{+\infty} \frac{d \omega}{\omega^2} e^{\mathrm{i} \omega t} \sum_{\ell m} {}_{-2}S_{\ell m}(\iota, \varphi_0) R_{\ell m n}(\omega,r)
    \end{equation}
    where the overtone index $n$ is now introduced to denote the eigen functions with different QNMs $\omega$, $\iota$ is the angle between the angular momentum vector and the line-of-sight vector while $\varphi_0$ is the phase angle based on source frame;~\cite{varma2014gravitational}. It should be noted that this formalism of reconstruction into the linear superposition of different modes is incompleteness, but the numerical simulation shows the applicability of this formalism at intermediate times;~\cite{berti2006gravitational,szpak2004quasinormal,beyer1999completeness,beyer2001stability}.
    
    Then, by separating the real and imaginary parts of QNMs $\omega=\omega_{\ell m n}+\frac{\mathrm{i} }{ \tau_{\ell m n}}$ and substituting the outgoing boundary behavior given from Eq.\eqref{TeuBCInf} as $R_{\ell m n}=r^3 Z_{\ell m \omega}^{\text {out }} e^{-\mathrm{i} \omega r}$ with $Z_{\ell m \omega}^{\text {out }}=M \mathcal{A}_{\ell m n} \mathrm{e}^{\mathrm{i} \phi_{\ell m n}}$, the polarization amplitudes can be obtained as:
    \begin{equation}
        h_{+}=\frac{M}{r} \operatorname{Re}\left[\mathcal{A}_{\ell m n}^{+} \mathrm{e}^{\mathrm{i}\left(\omega_{\ell m n} t+\phi_{\ell m n}^{+}\right)} \mathrm{e}^{-t / \tau_{\ell m n}} S_{\ell m n}(\iota, \varphi_0)\right]
    \end{equation}
    \begin{equation}
        h_{\times}=\frac{M}{r} \operatorname{Im}\left[\mathcal{A}_{\ell m n}^{\times} \mathrm{e}^{\mathrm{i}\left(\omega_{\ell m n} t+\phi_{\ell m n}^{\times}\right)} \mathrm{e}^{-t / \tau_{\ell m n}} S_{\ell m n}(\iota, \varphi_0)\right]
    \end{equation}
    where $\mathcal{A}_{\ell m n}^{+,\times}$ and $\phi_{\ell m n}^{+,\times}$ are the amplitude and the original phase respectively in general regarded as the free parameters or determined by the previous stage;~\cite{PhysRevD.90.124032,taracchini2012prototype}. Meanwhile, $\omega_{\ell m n}=2\pi f_{\ell m n}$ is the QNMs' real part with $f_{\ell m n}$ the frequency of the oscillation, and $\tau_{\ell m n}$ is damping time given from the values of QNMs;~\cite{berti2009quasinormal}. For the fundamental mode of Schwarzschild case with $m=0$, $l=2$ and $\omega=0.747343-0.177925\mathrm{i}$ under the unit $c=G=2M=1$, it turns out:
    \begin{equation}
        f_{200}=\pm 1.207 \cdot 10^{-2}\left(\frac{10^6 M_{\odot}}{M} \right) \mathrm{Hz}
    \end{equation}
    \begin{equation}
        \tau_{200}=55.37\left(\frac{M}{10^6 M_{\odot}} \right) \mathrm{s}
    \end{equation}
     In general, the ringdown waveform is dominated by the mode with $\ell=|m|=2$ for Kerr case, while the other multipoles are subdominant;~\cite{berti2007inspiral,buonanno2007inspiral}. And the QNMs for given $(\ell,m)$ are sorted by the damping time $\tau_{\ell m n}$, where the fundamental mode noted by $n=0$ has the longest damping time with the integer overtone index $n>0$ labeling the other modes with shorter damping time. And the mode with $(\ell, m, n)=(2,2,0)$ or noted as $(2,2,0)$ mode is usually the fundamental mode for Kerr case. Meanwhile the measurement of a detector is given by:
    \begin{equation}
        h=h_{+} F_{+}\left(\theta_S, \phi_S, \psi_S\right)+h_{\times} F_{\times}\left(\theta_S, \phi_S, \psi_S\right)
    \end{equation}
    with the pattern functions given as: 
    \begin{equation}
        \begin{aligned}
         F_{+}\left(\theta_S, \phi_S, \psi_S\right)=& \frac{1}{2}\left(1+\cos ^2 \theta_S\right) \cos 2 \phi_S \cos 2 \psi_S-\cos \theta_S \sin 2 \phi_S \sin 2 \psi_S
        \end{aligned}
    \end{equation}
    \begin{equation}
        \begin{aligned}
        F_{\times}\left(\theta_S, \phi_S, \psi_S\right)=& \frac{1}{2}\left(1+\cos ^2 \theta_S\right) \cos 2 \phi_S \sin 2 \psi_S +\cos \theta_S \sin 2 \phi_S \cos 2 \psi_S .
        \end{aligned}
    \end{equation}
    Where $\theta_S$ and $\phi_S$ denote the polar and azimuth angles of the source in the sky based on detector frame, while $\psi_S$ is the azimuth angles of the angular momentum vector based on radiation frame;~\cite{varma2014gravitational}. Besides the parameters of the source directly determining the waveform generally including the mass $M$ and the spin parameter $a$, another critical parameter associated to the detectability and measurability is the signal-to-noise ratio (SNR) $\rho$ defined as\cite{finn1992detection,flanagan1998measuring1,flanagan1998measuring2}:
    \begin{equation}
        \rho^2=4 \int_0^{\infty} \frac{\tilde{h}^*(f) \tilde{h}(f)}{S_h(f)} d f
    \end{equation}
    where $S_h(f)$ is the noise power spectral density (PSD) or sensitivity curve of different detectors;~\cite{LIGO,VIRGO,KAGRA,Robson_2019,PhysRevD.100.044042,wang2022hubble,PhysRevD.100.044042}. In general, SNR is the threshold to determine if a signal has been detected. When the SNR exceeds a certain threshold, such as $\rho>2.5$ in \cite{cabero2020black}, we consider the signal have been detected. Another method of applying the Bayesian model to examine the likelihood of detecting QNMs will yield the Bayes factor defined as:
    \begin{equation}
        \mathcal{B}_{A B}=\frac{p\left(d \mid H_A\right)}{p\left(d \mid H_B\right)}
    \end{equation}
    where $\mathcal{B}_{A B}>3.2$ denotes "substantial" support for $H_A$ over $H_B$, $\mathcal{B}_{A B}>10$ denotes "strong" support and $\mathcal{B}_{A B}>100$ is “decisive”;~\cite{kass1995bayes,cabero2020black}.
    
    Before the first detected GW event GW150914;~\cite{GW150914}, there have been several studies to predict the range of measurable sources, such as the results of \cite{flanagan1998measuring1}:
    \begin{equation}\label{QNMsRange}
        \begin{cases}
          60 M_{\odot} \lesssim M \lesssim 1000 M_{\odot}& \text{LIGO initial}  \\
            200 M_{\odot} \lesssim M \lesssim 3000 M_{\odot} & \text{Advanced LIGO}  \\
            10^7 M_{\odot} \lesssim M \lesssim 10^9 M_{\odot}& \text{LISA}
        \end{cases}
    \end{equation}
    As Earth-based GWs detectors continuously probe GWs, data-driven searches for the existence or the accurate detection of QNMs (or overtone) are taking a new direction. Meanwhile, the sources of some detected GWs events are just within the range predicted by Eq.\eqref{QNMsRange};~\cite{GWOSC,GWTC1,GWTC2,GWTC2.1,GWTC3}, and we summarize some recent advancements in this subject:
    
    \begin{itemize}
        \item \textbf{Detection of Fundamental Mode and Higher Overtone}. Using a waveform model with ringdown process has shown obvious advancements in the parameter estimates of mass $M$ and spin $a$ contrast with that without ringdown process;~\cite{baibhav2018black,PhysRevLett.123.111102}. However, due to the short damping time of the QNMs, it is difficult to identify if the QNMs included in the data and when they happened;~\cite{PhysRevD.90.124032,PhysRevD.94.069902}. Therefore, the standard QNMs tests only take the fundamental $(2,2,0)$ mode into consideration.
    
        However, the fundamental mode alone is not enough to estimate the accurate values of mass $M$ and spin $a$ because of lack of the information when the ringdown process happens. The research from \cite{giesler2019black} tried to consider a model including overtones up to $n=7$ and claimed that the spacetime can be well described as a linear perturbed BH directly after the peak. This means that extending the ringdown process directly after the peak by introducing higher overtones may be feasible, with the great significance for how to connect the ringdown waveform after the previous waveform and has led to a series of researches in \cite{PhysRevD.101.044033,PhysRevD.102.044053,PhysRevD.103.044054,PhysRevD.102.024027,PhysRevD.103.104048,finch2021modeling,PhysRevD.105.104015,jaramillo2022gravitational}.
    
        Meanwhile the revisiting for the first detected GW GW150914 brought a contradict problem. As is pointed in \cite{cotesta2022analysis}, both of the researches in \cite{PhysRevD.103.024041,PhysRevD.103.122002} provided a weak evidence in favor of "overtone model" with $\text{log}_{10}$-Bayes factor $\sim 0.6$, contradicting with the research in \cite{PhysRevLett.123.111102} who claimed at least one of overtones detected with $3.6\sigma$ confidence. As this problem is under further exploration, a startling new point appears that the overtones already detected may be noise-dominated because of the low Bayes factors! And some of the related researches can be found in \cite{cotesta2022analysis,isi2022revisiting}.
    
        In general, introducing overtones into waveform model indeed brings better effects. However, how to take them into a waveform model and if the overtones detected are noise-dominated may require further exploration and more accurate detection.
    
    \item \textbf{Detection of Higher Angular Modes}. In general, the $(2,2,0)$ mode is indeed the dominant mode, while the sub-dominant mode is sometimes not the $(2,2,1)$ mode but the modes with $\ell>2$ or $m>2$ called higher angular modes. And some related publications are \cite{capano2022statistical,dhani2021overtones,PhysRevD.102.044053,PhysRevD.102.044053,finch2021modeling,PhysRevD.105.104015}.
    
    \item \textbf{Detection of Nonlinear QNMs}. We have already known that under the inclusion of overtones, the waveform can be well-modelled. However, a binary BH merger is a highly nonlinear system, and we don't know if this nonlinear behavior propagates to infinity and contribute to the waveform on our detectors. In Sec.\ref{section:QNMeqn}, the QNMs are produced under the linear perturbations. When we take the second or higher order terms in Eq.\eqref{LinearMetric} into consideration, the nonlinear QNMs can be calculated with the details in \cite{brizuela2009complete} for Schwarzschild case and in \cite{loutrel2021second,ripley2021numerical} for Kerr. In the meanwhile, the current detectability research revealed that the detection of nonlinear QNMs requires more precise detectors, which may be available in the next generation;~\cite{cheung2022nonlinear,mitman2022nonlinearities}.
    
    \end{itemize}
   
\section{Discussion}
    
    We review the QNMs produced by linear perturbation theory in Sec.\ref{section:QNMeqn}, where we discuss the difficulties of reconstruction of metric and summarize some publications including the formalism of QNMs equations for tensor perturbations. The method to calculate the QNMs are reviewed in Sec.\ref{section:Method} including the newly developed methods in the past few decades. At the end of the article in Sec.\ref{Section:detection}, we review the detection advancements and highlight the current difficulties in detection of overtones.

\subsection{Tables}

    \begin{table}
        \centering
        \begin{tabular}{|c||c|c|c|}
	    	\hline
	    	\text{Perturbation Type} &\text{Scalar}&\text{Vector}&\text{Tensor} \\ 
	    	\hline
	    	$s$ & 0 & +1 or -1 & +2 or -2\\
	    	\hline
            $\psi$ & $\Phi$ & $\psi_0$ or $\tilde{\rho}^{-2}\psi_2$ & $\Psi_0$ or $\tilde{\rho}^{-4}\Psi_4$\\
	    	\hline
	    \end{tabular}%
	    \caption{
	    Table lists the specific form of Teukolsky function for different value of $s$ corresponds to the different type of perturbations in Kerr spacetime, where $\Phi$ is the wave function of the scalar master equation Eq.\eqref{CGeqn}, $\psi_0$ and $\psi_2$ are the Newman-Penrose (N-P) quantities of Maxwell field defined by Eq.\eqref{NPVector} while $\Psi_0$ and $\Psi_4$ describe gravitational radiation defined by Eq.\eqref{NPTensor}. And for Kerr case, $\tilde{\rho}=-\frac{1}{r-\mathrm{i} a \cos \theta}$ is a variable in N-P formalism as defined in Eq.\eqref{NPrho}.
	    }
	    \label{Table:TeuFunction}
    \end{table}

\section*{Figure captions}


\setcounter{figure}{2}
\setcounter{subfigure}{0}
\begin{subfigure}
\setcounter{figure}{2}
\setcounter{subfigure}{0}
\centering
    \begin{minipage}[b]{0.5\textwidth}
        \includegraphics[width=\linewidth]{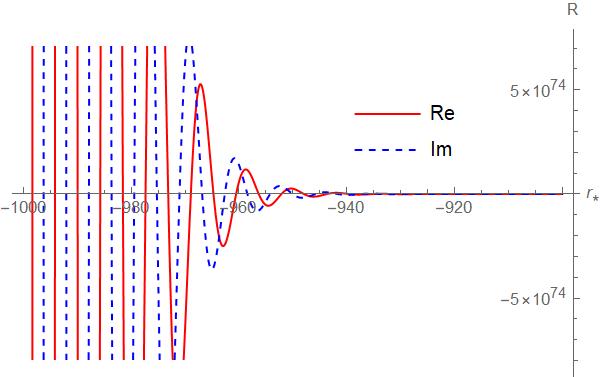}
        \caption{behavior near horizon under tortoise coordinate}
        \label{fig:SchBCH}
    \end{minipage}  
   
\setcounter{figure}{2}
\setcounter{subfigure}{1}
    \begin{minipage}[b]{0.5\textwidth}
        \includegraphics[width=\linewidth]{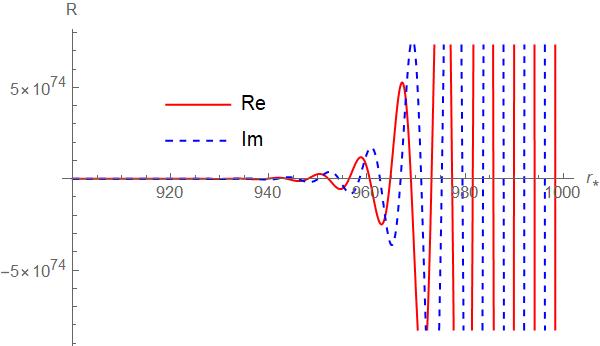}
        \caption{behavior at infinity under tortoise coordinate}
        \label{fig:SchBCInf}
    \end{minipage}

\setcounter{figure}{2}
\setcounter{subfigure}{-1}
    \caption{Fig.\ref{fig:SchBCH} describes the exponentially diverging asymptotic behavior near horizon under the tortoise coordinate while Fig.\ref{fig:SchBCInf} pictures that of spatial infinity, where we use the unit c=G=2M=1 and the fundamental one of QNMs in Schwarzschild case with $\ell=2$ and $\omega=0.747343-0.177925\mathrm{i}$.}
    \label{fig:SchBC}
\end{subfigure}

\begin{figure}[h!]
    \begin{center}
    \includegraphics[width=10cm]{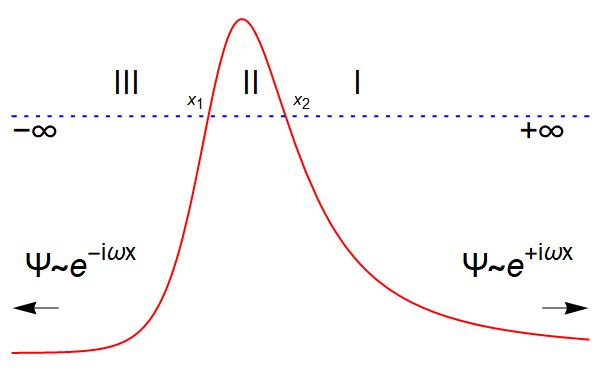}
    \end{center}
    \caption{This is the schematic diagram of WKB method where two turning points $x_1$ and $x_2$ are determined by $\omega^2\sim V_{\text{eff}}$, which make the whole integration domain divided into three regions.}\label{fig:WKB}
\end{figure}

\bibliographystyle{Frontiers-Vancouver}
\bibliography{reference}

\end{document}